\documentclass[12pt]{article}
\UseRawInputEncoding
\begin{document}

\begin{center}
{\large \bf Consequences of the MSW mechanism with Super-Kamiokande oscillation parameters and of the volume distribution of neutrino sources in the Sun}
\vspace{0.5 cm}

\begin{small}
\renewcommand{\thefootnote}{*}
L.M.Slad\footnote{slad@theory.sinp.msu.ru} \\
{\it Skobeltsyn Institute of Nuclear Physics,
Lomonosov Moscow State University, Moscow 119991, Russia}
\end{small}
\end{center}

\vspace{0.3 cm}

\begin{footnotesize}
\noindent
{\bf Abstract.} The knowledge of the parameters of the solar neutrino oscillation model, provided by the SNO and Super-Kamiokande collaborations, enables us to obtain, with the Wolfenstein equation, a simple and clear analytical and numerical picture of the transformation of the neutrino state during its travel inside the Sun. We show that the resulting picture is not implicated in the characteristics of the neutrino state at the surface of the Sun and at the surface of the Earth. This circumstance with taking into account the volume distribution of solar neutrino sources indicates an obvious contradiction between the consequences of the MSW mechanism with parameters from SNO and Super-Kamiokande and the results of three of the five observed processes with solar neutrinos. Erroneous assertions at the base of the MSW mechanism are noted.  

\vspace{0.5 cm}
\noindent
{\bf Keywords:} differential equations; solar neutrino sources; electron neutrino survival

\end{footnotesize}

\vspace{0.5 cm}

\begin{small}

\begin{center}
{\large \bf 1. Introduction}
\end{center}

The present work is devoted to a logically simple analysis of a number of aspects of the Mikheev-Smirnov-Wolfenstein (MSW) mechanism \cite{1}, \cite{2} that remained in the shadow, as well as to taking into account the volume distribution of neutrino sources in the Sun. This mechanism, according to the widespread opinion of the physical society, provides a solution to the solar neutrino problem based on transitions in matter of electron neutrinos in neutrinos of other types.  The need for such an analysis is due to several circumstances.

First, an elegant alternative solution to the solar neutrino problem based on a hypothesis of the existence of a new interaction has been appeared \cite{3}. The carrier of this interaction is the postulated massless pseudoscalar boson, which has a Yukawa coupling to, at least, electron neutrinos and nucleons but not to electrons. Collisions of neutrinos with nucleons of the Sun lead to almost equal fluxes of left- and right-handed electron neutrinos at the Earth's surface and to a decrease in the energy of these neutrinos. Having a single free parameter, this hypothesis gives good agreement between the theoretical and experimental characteristics for all five observed processes with solar neutrinos.

Second, the results of experiments with solar neutrinos in the Sudbury Neutrino Observatory (SNO) \cite{4} and in the Super-Kamiokande \cite{5} are interpreted on the basis of the neutrino oscillation model with well-defined values ​​of its parameters. This makes it possible to narrow the arbitrariness in the MSW mechanism, to carefully consider a number of essential aspects of the realization of this mechanism and to give a new assessment of its capabilities. Numerical results obtained after analytical calculations show a simple and clear picture of the transformation of the state of a neutrino during its travel inside the Sun and point to the fact that this picture is not implicated in the characteristics of the neutrino state at the surface of the Sun and at the surface of the Earth.

All the starting points of the Mikheev-Smirnov-Wolfenstein mechanism concerning transformations of a neutrino when propagating in matter are fully formulated in the work of Wolfenstein \cite{1}. In this paper, two hypothetical variants of neutrino oscillations are proposed. In the one variant, a four-fermion interaction model with neutral neutrino V-A current containing terms that change the type of neutrino is considered. In the another variant, the neutral neutrino current is given by the standard model. Confining himself to two types of flavour neutrinos, $\nu_{e}$ and $\nu_{\mu}$, Wolfenstein assumes the existence of two more neutrinos $\nu_{1}$ and $\nu_{2}$ with unequal masses $m_{1}$ and $m_{2}$ ($m_{1} > m_{2}$), whose states in vacuum are given by the relations
\begin{equation}
|\nu_{1}\rangle = |\nu_{e}\rangle \cos \theta - |\nu_{\mu}\rangle \sin \theta,
\label{1}
\end{equation}
\begin{equation}
|\nu_{2}\rangle = |\nu_{e}\rangle \sin \theta + |\nu_{\mu}\rangle \cos \theta.
\label{2}
\end{equation}
Mikheev and Smirnov, in their various interpretations of neutrino oscillations in matter \cite{2}, rely on all of the initial provisions of Wolfenstein \cite{1}, except the variant with off-diagonal neutral neutrino currents.

We also restrict ourselves to the second mentioned variant and give an analytical and numerical analysis of the consequences of the Wolfenstein equation related to solar neutrinos and presented as (see \cite{6}, Eq. (14.56))
\begin{eqnarray}
& &i\frac{d \;}{dt}\left(
\begin{array}{c}
A_{e}(t) \\
A_{\mu}(t)
\end{array} 
\right) = \frac{1}{2}\left(
\begin{array}{cc}
\displaystyle -\frac{\Delta m^{2}}{2E} \cos 2\theta + \sqrt{2} G_{F} N_{e}(t) & \displaystyle \frac{\Delta m^{2}}{2E} \sin 2\theta \\
\displaystyle \frac{\Delta m^{2}}{2E} \sin 2\theta & \displaystyle \frac{\Delta m^{2}}{2E} \cos 2\theta - \sqrt{2} G_{F} N_{e}(t)
\end{array}
\right) \times \nonumber \\
& & \times \left(
\begin{array}{c}
A_{e}(t) \\
A_{\mu}(t)
\end{array}
\right), \label{3}
\end{eqnarray}
where $A_{e}(t)$ ($A_{\mu}(t)$) is the probability amplitude for the neutrino state to be electronic (muonic) type at the time moment $t$, when the electron density at the neutrino location is equal to $N_{e}(t)$; $\Delta m^{2} = m_{1}^{2}- m_{2}^{2}$; $E$ is the neutrino energy.

We rely on a number of numerical results of the standard solar model (SSM) contained in the review \cite{7}: on the dependence of the matter density of the Sun on the distance to its center, on the distribution of neutrino sources over the volume of the Sun, and on the boundary values of the energy of solar neutrinos. In our numerical analysis, we use the central values of the parameters of the two-neutrino model of solar neutrino oscillations shown in the most recent paper by the Super-Kamiokande collaboration \cite{5}:
\begin{equation}
\Delta m^{2} = 4.8^{+1.5}_{-0.8} \times 10^{-5} \; {\rm eV}^{2}, 
\label{4}
\end{equation}
\begin{equation}
\sin^{2} \theta = 0.334^{+0.027}_{-0.023} .
\label{5}
\end{equation}

\begin{center}
{\large \bf 2. Neutrino at the surface of the Sun after their transformation of its state when moving inside the Sun}
\end{center}

Adhering to the concept of the continuity of physical quantities, we attribute the Wolfenstein equation (\ref{3}) to the entire time interval from the moment of the neutrino production in the Sun to its registration on the Earth, while the electron density at a neutrino’s location can be quite large or arbitrarily small and zero values. To deny such continuity would be unprecedented in physics. Thus, the Mikheev-Smirnov-Wolfenstein mechanism must apply to neutrinos moving both, inside the Sun and in the vacuum outside the Sun where it reduces to the standard oscillation model.

We suppose that the electron density $N_{e}(t)$ in the Sun during several oscillations in the time interval from $t_{0}$ to $t$ can be considered constant with sufficient accuracy: $N_{e}(t) = N_{e}(t_{0}) \equiv N_{e}$. In what follows, we will make sure that the corresponding relative change in the electron density is approximately $10^{- 3}$.

Then, Eq. (\ref{3}) reduces to a linear homogeneous differential equation of the second degree with constant coefficients with respect to the probability amplitude $A_{e}(t)$. We require that the solution for $A_{e}(t)$ and for $A_{\mu}(t)$ obtainable from it satisfy the following initial conditions
\begin{equation}
A_{\mu}(t_{0}) = 0, \qquad |A_{e}(t_{0})|^{2} = 1,
\label{6}
\end{equation}
showing that at the time moment $t_{0}$ the neutrino state is purely electronic. The respective probability $P_{e}(t)$ that the neutrino state at the time moment $t$ is electronic is given by the formula
\begin{eqnarray}
& &P_{e}(t) = |A_{e}(t)|^{2} = \frac{1}{2} \left( 1+\frac{1}{f(N_{e})^{2}}\left( \left( \frac{\Delta m^{2}}{2E}\right) \cos 2\theta -\sqrt{2} G_{F} N_{e} \right)^{2} \right) + \nonumber \\
& &+ \frac{1}{2} \left( 1-\frac{1}{f(N_{e})^{2}}\left( \left( \frac{\Delta m^{2}}{2E}\right) \cos 2\theta -\sqrt{2} G_{F} N_{e} \right)^{2} \right) \cos f(N_{e})(t-t_{0}), 
\label{7}
\end{eqnarray}
where
\begin{equation}
f(N_{e}) = \sqrt{\left( \frac{\Delta m^{2}}{2E}\right) ^{2} -2\sqrt{2}\left( \frac{
\Delta m^{2}}{2E} \right) G_{F} N_{e} \cos 2\theta + 2 G_{F}^{2} N_{e}^{2}}.
\label{8}
\end{equation}

It follows from Eq. (\ref{7}) that the oscillation length of the electron neutrino is
\begin{equation}
L(N_{e})/c = 2\pi/f(N_{e}).
\label{9}
\end{equation}
Taking into account Eq. (\ref{8}), we find that this length as a function of the electron density $N_{e}$ has a maximum at the density
\begin{equation}
(N_{e})_{\rm ext} = \frac{\cos 2\theta}{\sqrt{2} G_{F}} \left( \frac{\Delta m^{2}}{2E}\right), 
\label{10}
\end{equation}
at that
\begin{equation}
L(N_{e})_{\rm max}/c = \frac{2\pi}{\sin 2 \theta} \left( \frac{2E}{\Delta m^{2}}\right).
\label{11}
\end{equation}

At an extremal electron density (\ref{10}), the probability $P_{e}(t)$ can take any value from 1 to 0. In other cases, the probability value $P_{e}(t)$ belongs to the interval from 1 to $a$, $ 0 <a <1 $.

It follows from Eq. (\ref{8}) that the oscillation length of an electron neutrino in vacuum is given by the standard formula
\begin{equation}
L(0)/c = 2\pi \left( \frac{2E}{\Delta m^{2}}\right).
\label{12}
\end{equation}

We turn now to the numbers.

We note first of all that the number of electrons per one nucleon of the Sun $Y_{e}$ increases with increasing distance from the center of the Sun by about a factor of 1.3 \cite{7}. For definiteness, which does not affect our conclusions in any way, we assume that the value of $Y_{e}$ stays the same everywhere in the Sun,
$Y_{e} = 0.86$, which corresponds to mass fractions of hydrogen and helium equal to 0.74 and 0.25, respectively \cite{8}. Then the relation between the electron density $N_{e}$ and the matter density in the Sun $\rho$ is given by
\begin{equation}
N_{e} = 5.14 \cdot 10^{23} \left( \frac{\rho}{{\rm g\cdot cm}^{-3}} \right) {\rm cm}^{-3}.
\label{13}
\end{equation}

It follows from here that the maximum value of $N_{e}$ corresponding to the maximum value of the matter density in the center of the Sun $\rho = 148 \;{\rm g\cdot cm}^{-3}$ is
\begin{equation}
(N_{e})_{\rm max} = 7.61 \cdot 10^{25} \;{\rm cm}^{-3}.
\label{14}
\end{equation}

We obtain from Eqs. (\ref{4}), (\ref{5}), (\ref{10}), and (\ref{14})
\begin{equation}
(N_{e})_{\rm ext} > (N_{e})_{\rm max}, \qquad {\rm if} \;\; E < 0.827 \;{\rm MeV},
\label{15}
\end{equation}
i.e. for solar neutrinos having the energy $E < 0.827$ MeV, including all neutrinos from $p-p$ interactions, the oscillation length of their states has no extreme. It monotonically decreases with the decreasing the matter density of the Sun in the neutrino location.

We obtain from here and from Eqs. (\ref{4}), (\ref{5}), (\ref{8}), (\ref{9}), (\ref{12}), and (\ref{14}) that at a neutrino energy $E = 0.233$ MeV, which is the threshold energy for the transitions $\nu_{e}+{}^{71}{\rm Ga} \rightarrow e^{-}+{}^{71}{\rm Ge}$, the length of the probability oscillations $P_{e}(t)$ lies in the interval from 12.0 km to 12.4 km, i.e. from $0.0000172 R_{0} $ to $0.0000178 R_{0}$, where $R_{0}$ is the radius of the Sun, $R_{0} = 696,000$ km. At a neutrino energy of $E = 0.827$ MeV, the extremal electron density in the Sun coincides with the maximum possible value, and the oscillation length of the probability $P_{e}(t)$ lies in the interval from 42.6 to 45.3 km, i.e. from $0.0000612 R_ {0}$ to $0.0000651 R_{0}$.

Consider the interval of solar neutrinos energies from 0.827 MeV to the maximum possible 18.8 MeV, that corresponds to neutrinos from $hep$. For a fixed energy from this interval, the oscillation length of the probability $P_{e}(t)$ has a maximum value at the matter density of the Sun corresponding to the extremal electron density (\ref{10}). When the neutrino energy increases from 0.827 to 18.8 MeV, the extremal density of matter decreases from 148 to 6.51 g/cm$^{3}$. The oscillation length of the probability $P_{e}(t)$ takes a minimum value at one of the boundary points of the matter density range [0, 148] g/cm$^{3}$. At the neutrino energy $E = 18.8$ MeV, the oscillation length of the probability $P_{e}(t)$ lies in the interval from 134 km to 1028 km, i.e. from $0.000193 R_ {0}$ to $0.00148 R_{0}$, at that it is equal $L(0) = 968$ km at the exit from the Sun.

Using the dependence of the matter density in the Sun on the distance from its center, presented in \cite {7}, we are convinced that the change in the matter density during the maximum period of oscillations of the neutrino state in the Sun, which corresponds to the oscillation length $0.00148 R_{0}$, is very extremely small: $\Delta \rho = 1.7 \cdot 10^{-3} \rho$. Consequently, the accepted condition about the constancy of the electron density during
any admissible period of oscillations is well justified.

At the time moment $t_{1} = t_{0}+L(N_{e})/c$, we have, as it follows from Eqs. (\ref{7}) and (\ref {9}), that the equality $P(t_ {1}) = 1$ is right, this is, if the neutrino state was purely electronic at the time moment $t_{0}$ in some place of the Sun, it becomes again purely electronic at the time moment $t_{1}$ in the corresponding new place of the Sun. Therefore, we can turn to the Wolfenstein equation (\ref{3}) again to describe the transformation of the state of the solar neutrino, starting in the time moment $t_{1}$, for what it is enough to replace the time moment $t_{0}$ in the previous formulas with $t_{1}$. At that, it is extremely important to note that this step of the description of the neutrino motion with unchanged energy and momentum direction is entirely characterized by the location of neutrino at the time moment $t_{1}$ and by the electron density $N(t_{1})$ in that location. It does not carry any information on the location of neutrino at the time moment $t_{0}$ and about the corresponding electron density $N(t_{0})$, as well as on the neutrino state transformation between $t_{0}$ and $t_{1}$.

We can proceed, step by step, to new time moments $t_{n}$ of completing the next oscillations of the solar neutrino state and restoring it as purely electronic, when $P(t_{n}) = 1$, until neutrino at the time moment $t_{n_{S}}$ approaches the surface of the Sun by a distance less than the oscillation length $L(0)$ (\ref {12}) corresponding to the neutrino energy $E$. At that, the description of the neutrino state transformation after that time moment with the Wolfenstein equation (\ref{3}) neither refers to the neutrino production place, nor to the trajectory of its motion, nor to the nature of the change in the oscillation length along this trajectory. From the time moment $t_{n_{S}}$ on, the transformation of the state of the solar neutrino on its way to the Earth is described by the standard model of oscillations in vacuum. The probability for a solar neutrino to be electronic type when reaching the experimental setup on Earth at the time moment $t_{E}$ is given, as it follows from Eqs. (\ref{7} and (\ref{8}), by 
\begin{equation}
P_{e}(t_{E}) = \frac{1}{2}(1+\cos^{2} 2\theta)+\frac{1}{2}(1-\cos^{2} 2\theta)\cos (\Delta m^{2}/2E)(t_{E}-t_{n_{S}}).
\label{16}
\end{equation}

\begin{center}
{\large \bf 3. Volume distribution of solar neutrino sources and averaging over the neutrino states at the Earth}
\end{center}

According to the standard solar model, each neutrino source $s$ has a sufficiently wide spherically symmetric distribution over the solar volume \cite{7} (Fig. 8). So, for neutrinos from the decays of $^{8}$B, the distribution reaches its maximum at a distance of $0.045 R_{0} = 31,000$ km from the center of the Sun, and its width at half amplitude level is $0.053 R_{0} = 37,000$ km. For neutrinos from $p-p$ collisions, the respective parameters are $0.103 R_{0} = 72,000$ km and $0.11 R_{0} = 77,000$ km.

This fact is undoubtedly known to many proponents of the neutrino oscillation concept. For example, it is mentioned in \cite{2}. However, it is not reflected in the periodically updated review of this concept \cite{6}.

The mentioned individual numerical characteristics of the distribution of neutrino sources indicate that almost parallel neutrino fluxes, generated in different places of the Sun and entering an experimental setup on Earth, can differ in the length of their trajectories, both inside the Sun and in vacuum, by tens and hundreds of thousands of kilometers. Since the maximum oscillation length inside the Sun and in the vacuum (at $E = 18.8$ MeV) is approximately 1000 km, the difference in the numbers of neutrino oscillations along the various mentioned trajectories can be several tens or hundreds. The variability of trajectories leads to the fact that the values of the cosine in Eq. (\ref{16}) cover the entire interval from -1 to 1. The electron neutrino flux at the Earth’s surface coming from the source $s$, $\Phi_{e}(s)$, is found by summing over the neutrino fluxes along various trajectories multiplied by the corresponding probabilities
\begin{equation}
\Phi_{e}(s) = \int P_{e}(t_{E})d\Phi(s) = \overline{P_{e}(t_{E})} \int d\Phi(s) =  \overline{P_{e}(t_{E})}\Phi(s), 
\label{17}
\end{equation}
where $\Phi(s)$ is the neutrino flux from the source $s$ given by the SSM. For the average probability $\overline{P_{e}(t_{E})}$, we take the result of Eq. (\ref{16}) averaged over the cosine argument in the range from 0 to $2 \pi$.

As a result we obtain for solar electron neutrinos survival probability $P_{ee}$ at the Earth's surface a well known expression
\begin{equation}
P_{ee} \equiv \overline{P_{e}(t_{E})} = \frac{1}{2}(1+\cos^{2} 2\theta) = 1-\frac{1}{2}\sin^{2} 2\theta, 
\label{18}
\end{equation}
which gives, taken together with (\ref{5}),
\begin{equation}
P_{ee} = 0.555 \pm 0.018.
\label{19}
\end{equation}

The probability $P_{ee}$ (\ref{19}) obviously contradicts the fact that, in three of the five observed processes with solar neutrinos, the ratio ${\cal K}$ of every experimental rate to the theoretical one calculated in the framework of the SSM is no more than 0.5. So, omitting orders in the values of that or another rate and giving first a reference to the experimental work, and then to the theoretical one, we have: for the nuclear transitions $^{37}{\rm Cl} \rightarrow ^{37}{\rm Ar}$ the ratio ${\cal K}$ is ($2.56 \pm 0.24$)/($7.9 \pm 2.6$) = $0.32 \pm 0.14$ [9] vs [7]; for the elastic scattering of solar neutrinos on electrons $\nu_{e}+e^{-} \rightarrow \nu_{e}+e^{-}$ the ratio ${\cal K}$ is ($2.32 \pm 0.07$)/($5.79 \times (1 \pm 0.23)$) = $0.40 \pm 0.10$ [10] vs [11]; for the deuteron disintegration by the neutral currents $\nu_{e}+D \rightarrow \nu_{e}+n+p$ the ratio ${\cal K}$ is ($1.76 \pm 0.11$)/($5.79 \times (1 \pm 0.23)$) = $0.30 \pm 0.09$ [12] vs [11].

For the nuclear transitions $^{71}{\rm Ga} \rightarrow ^{71}{\rm Ge}$ the ratio ${\cal K}$ within the error does not contradict the value in Eq. (\ref{19}). Namely, ${\cal K}$ is ($65.4 \pm 2.9$)/($131 \pm 10$) = $0.50 \pm 0.06$ [13] vs [11]. The experimental and theoretical rates of the process of deuteron disintegration by neutral currents $\nu_{e}+D \rightarrow \nu_{e}+n+p $ are close to each other. 

The three-neutrino variant of solar neutrino oscillations comprising additional oscillations with  short length and small amplitude, as it is the case of  interpretating a number of experiments with antineutrinos from nearby reactors (see, for example, \cite{14}), does not change the above conclusion. Our reasoning remains valid if we change the value of $\Delta m^{2}$ in equality (\ref{4}) by an order of magnitude and the value of $\sin^{2} \theta$ in equality (\ref{5}) by a factor of 2.

\begin{center}
{\large \bf 4. Concluding remarks}
\end{center}

The key point of this paper is the established fact that the characteristic sizes of the distributions of neutrino sources in the volume of the Sun are thousands of times greater than the maximum oscillation lengths of solar neutrinos at any values of their energies. All that we aimed to obtain within the framework of the considered model of solar neutrino oscillations, based on the Wolfenstein equation with Super-Kamiokande oscillation parameters, is to find the set of permissible oscillation lengths inside the Sun. The details of the transformation of the state of a solar neutrino when it moves along any trajectory do not affect the survival value of electron neutrinos near the Earth's surface.
 
\begin{center}
﻿{\large \bf 5. An addition}
\end{center}

   In the initial version of this article, I have avoided discussing the starting statements, reasoning and conclusions contained in any work related to the MSW mechanism.

   Experts informed in this version believe that the author must explain why the assertion about the adiabatic flavor conversion of solar neutrinos is wrong. I fulfill this wish.

   Hereinafter, I restrict my analysis only to certain aspects of the article by Mikheev and Smirnov \cite{2}, using its Russian-language version.

   I emphasize again that the only analytical and logical tool available to the MSW mechanism is the Wolfelstein equation (\ref{1}) written in the basis $(\nu_{e}, \nu_{\mu})$ or in the basis $(\nu_{1}, \nu_{2})$. 

   The authors of \cite{2} undoubtedly keep in mind the facts that, firstly, the experimental rate of transition of chlorine to argon under the action of solar neutrinos is approximately 1/3 of the theoretical rate found in the framework of the standard solar model, and that, second, in conventional situations, the probability that the initial electron neutrino will remain electronic, averaged over the time of one oscillation, is not less than 1/2. Their construction is aimed at making it plausible that a situation arises in which the initial electron neutrino, turning out to be muonic at some time, subsequently oscillate as the initial muonic neutrino, i.e. the electron neutrino and muon neutrino are swapped. 

   As it follows from the words of Smirnov, said in \cite{15} (p. 16), he well understands the role of terminology in the perception of his construction of solar neutrino transformations: "terminology should reflect and follow our understanding of the subject". The role of the terms "resonance" and "adiabaticity", inappropriate in elementary particle physics and field theory and used by Mikheev and Smirnov in their article \cite{2}, consists in creating the basis for the conclusions they planned. Let us look at the features of the transformations of neutrino states with which these terms are associated, comparing them with the solutions of the Wolfenstein equation for separate parts of the trajectories of solar neutrinos, the variety of which is described above in Section 2.
 
   If, above in Section 2, special attention was paid to the extremum of the oscillation length, then in [2] it was given to the extremum of the neutrino mixing angle.

   When moving in a medium with an electron density $N_{e}$, the wave functions of massive particles, now denoted as $\nu_{1m}$ and $\nu_{2m}$, differ from their wave functions in vacuum $\nu_{1}$ and $\nu_{2}$. Decomposition similar to (\ref{1}) and (\ref{2}) is written as 
\begin{equation}
|\nu_{1m}\rangle = |\nu_{e}\rangle \cos \theta_{m} - |\nu_{\mu}\rangle \sin \theta_{m},
\label{20}
\end{equation}
\begin{equation}
|\nu_{2m}\rangle = |\nu_{e}\rangle \sin \theta_{m} + |\nu_{\mu}\rangle \cos \theta_{m}.
\label{21}
\end{equation} 

   Thereof it follows that the neutrino state, which was purely electronic at the initial time $t_{0}$, remains electronic at the subsequent time $t$ with the probability 
\begin{equation}
P_{e}(t) = 1 - \sin^{2} \theta_{m} \sin^{2} \left( \frac{\pi (t-t_{0})c}{L(N_{e})}\right).
\label{22}
\end{equation}

   Representing the formula (\ref{7}) in the form (\ref{22}), we obtain 
\begin{equation}
\sin \theta_{m} = \frac{\Delta m^{2}}{2E} \cdot \frac{\sin 2\theta}{f(N_{e})},
\label{23}
\end{equation}
that reproduces formulas (1) and (2) of the work \cite{2}.
   
   On the part of the neutrino trajectory in the Sun, where the electron density is given by the formula (\ref {10}), and the oscillation length has an extremal value (\ref{11}), the quantity $\sin \theta_{m}$ also has an extremal value equal to the number of 1. Since $\sin \theta_{m}$ is a continuous function of the density of the Sun's matter, the change in this function is very little, 
\begin{equation}
\Delta \sin \theta_{m} = \tan^{-2} 2\theta \frac{\Delta \rho}{\rho},
\label{24}
\end{equation}
during oscillations with extreme values of a number of quantities. Taking into account Eq. (\ref{5}) and the fact noted in Section 2 that
$\Delta \rho / \rho < 1.7\cdot 10^{-3}$, we have
\begin{equation}
\Delta \sin \theta_{m} < 2.1 \cdot 10^{-4}.
\label{25}
\end{equation}

   Mikheev and Smirnov assert that the mixing angle in matter $\theta_{m}$ changes by $\pi /2$ during oscillations with extreme values ​​of a number of quantities. It is erroneous, since it contradicts the relation (\ref{25}), which is a consequence of the Wolfenstein equation (\ref{3}) when the real distribution of matter in the Sun is taken into account \cite{7}. Their statement is also erroneous that the initial beam of electron neutrinos, having passed a part of the trajectory with an extremal value ​​of the quantity $\sin \theta_{m}$, is almost completely transformed into a beam of muon neutrinos and at the exit from the Sun $P_{e} \simeq \sin^{2} \theta$, since it contradicts the relation (\ref{22}). Indeed, if at the instant $t_{0}$ the neutrino state was purely electronic, then at the instant $t_{c} = t_{0} + L(N_{e}) / 2c$ the neutrino state becomes purely muonic. For an arbitrarily small time interval that has elapsed after this moment, the nonzero probability of the electron state of the neutrino is restored. At the moment of time $t_{1} = t_{0} + L(N_{e}) / c$, the neutrino state again becomes purely electronic. 

   Recognition of the above statements of Mikheev and Smirnov to be true is equivalent to the denial of the consequences of the Wolfenstein equation (\ref{3}) for at least during the time of one oscillation with the maximum length, which leads, in particular, to the denial of the existence of an time moment when the neutrino state is purely muonic.  

  The opinion of Mikheev and Smirnov that in the state of a neutrino, after passing through a point with extrema of a number of quantities, the electronic and muonic components change places, has no logical and analytical justification and seems to be psychological fornication. 
  
   So, the statement about the existence of adiabatic flavor conversion of solar neutrinos, which serves as the basis of the MSW mechanism, is erroneous.

\end{small}
\end{document}